# Two-gap superconductivity in the noncentrosymmetric La$_3$Se$_4$ compound

F. Košuth,[1,2] N. Potomová,[2] Z. Pribulová,[1] J. Kačmarčík,[1] M. Naskar,[3] D.S. Inosov,[3] S. Ash,[4] A.K. Ganguli,[5,6] J. Šoltýs,[7] V. Cambel,[7] P. Szabó,[1] and P. Samuely[1]

[1]*Centre of Low Temperature Physics, Institute of Experimental Physics, Slovak Academy of Sciences, SK-04001 Košice, Slovakia*

[2]*Centre of Low Temperature Physics, Faculty of Science, P. J. Šafárik University, SK-04001 Košice, Slovakia*

[3]*Institut für Festkörper- und Materialphysik, Technische Universität Dresden, D-01069 Dresden, Germany*

[4]*Institute for Solid State Research, Leibniz IFW Dresden, 01069 Dresden, Germany*

[5]*Department of Chemistry, Indian Institute of Technology Delhi, New Delhi 110016, India*

[6]*Department of Chemical Sciences, Indian Institute of Science Education and Research Berhampur, Odisha 760003, India*

[7]*Institute of Electrical Engineering, Slovak Academy of Sciences, Bratislava, Slovakia*

Point-contact Andreev reflection spectroscopy at low temperatures and high magnetic fields has been performed on a noncentrosymmetric La$_3$Se$_4$ superconductor with a critical temperature $T_c$ = 8 K. Two superconducting energy gaps $\Delta_1$ and $\Delta_2$, with $2\Delta_1/k_BT_c$ ~ 5.8 and $2\Delta_2/k_BT_c$ ~ 2.3, are directly observed in some of the spectra. The temperature and magnetic field effects help to resolve a two-gap structure even on the most frequent spectra where at low temperatures only a single gap is apparent, reflected in a pair of maxima around the zero bias. Two-gap superconductivity consistently with the point contact Andreev reflection spectroscopy is also supported by the heat capacity and the Hall probe magnetization measurements.

*Point-contact Andreev reflection spectroscopy, heat capacity, Hall-probe magnetometry, penetration field, non-centrosymmetric superconductors, multiband/multigap superconductivity*

## 1 Introduction

Superconductors without inversion symmetry attract attention for their potential unique physical properties and possible applications. The absence of inversion symmetry combined with antisymmetric spin-orbit coupling lifting the Kramers degeneracy in these materials can lead to unconventional superconducting behavior, exotic pairing mechanisms with mixed parity, and novel phenomena such as topologically protected surface states and resilience to extreme external magnetic fields [1]. First heavy-fermion superconductor CePt$_3$Si [2] has been a primary candidate for exotic pairing mediated by magnetic fluctuations where a mixed-parity state yields different gap structures on the two spin-split Fermi surfaces, one with line nodes [3,4]. Antisymmetric spin-orbit interaction may cause multiple superconducting gaps on multiple bands crossing the Fermi level even without a triplet component. For example, in non-centrosymmetric sesquicarbides of La$_2$C$_3$ and Y$_2$C$_3$ two superconducting gaps of the s-wave symmetry have been detected [5], but also in LaNiC$_2$ [6]. The upper critical magnetic field exceeding the Pauli limit has been frequently observed in the noncentrosymmetric superconductors. Obviously, it is expected for the systems with a triplet component of the order parameter but the strong coupling effects with $\Delta > \Delta_{BCS}$ maybe at play, too.

Here, we address for the first time the problem of the superconducting gap in the noncentrosymmetric compound La$_3$Se$_4$. Superconductivity in multiband systems such as La$_3$Se$_4$ can feature different superconducting energy gaps on different bands, and strong spin-orbit coupling can bring a mixing of spin singlet and triplet states. Our point-contact Andreev reflection spectroscopy measurements supported by the low-temperature heat capacity and the Hall probe magnetization suggest the presence of two superconducting energy gaps in the system.

## 2 Experimental

Polycrystalline samples with a nominal composition of La$_3$Se$_4$ were synthesized by reacting La and Se (in elemental form). The superconducting transition temperature of this system which crystalizes in bcc structure of Th$_3$P$_4$ type is about 8 K. Details of the synthesis can be found in [7] together with the electrical transport, magnetic, and thermal properties of the system.

Point-contact Andreev reflection (PCAR) spectroscopy measurements were performed in a special point-contact approaching system enabling measurements down to $T = 1.5$ K and in magnetic fields up to $B = 8$ T [8]. Before the experiment, the samples typically a few millimeters in size were cut from larger $La_3Se_4$ polycrystalline pieces and then polished by $Al_2O_3$ sandpaper at ambient conditions. The point-contact microconstrictions were formed in situ by pressing a metallic tip to the polished surface of the sample using a differential screw mechanism. The standard lock-in modulation method was used to measure the differential resistance as a function of the applied voltage and subsequently, the differential conductance $\sigma(V) = dI(V)/dV$ was calculated (PCAR spectra).

The charge transfer at the ballistic point-contact microconstriction (PC) between a normal metal and a superconductor (N-S) at energies $eV$ smaller than the energy gap of the superconductor is realized through Andreev reflection. Since electrons with $|eV| < \Delta$ cannot pass directly through the PC, the transport is realized by the formation of Cooper pairs and the incident electrons retro-reflect holes from the PC interface. This leads to a doubling of the PC conductance at energies $|eV| < \Delta$. In real PCAR experiments, we usually observe a dip at zero bias accompanied by two distinct maxima at voltages close to that of the superconducting energy gap, $\Delta/e$, where $e$ is the electron's charge. This dip is a sign of the presence of a contact barrier. Upon increasing the contact barrier, the dip becomes deeper, and at a sufficiently strong barrier, the PC spectrum will have a pure tunneling character, with zero conductance at energies smaller than $\Delta$. The transport through a N-S PC can be described with the Blonder-Tinkham-Klapwijk model (BTK) [9], which introduces the superconducting energy gap $\Delta$, spectral smearing $\Gamma$ [10] and the PC barrier strength Z, and describes the evolution of the PC spectrum between a metallic ($Z = 0$) and a pure tunneling-like ($Z > 1$) case.

In the case of two-gap superconductors the PC spectrum typically reveals a double superconducting gap structure, which can be described as the weighted sum of two BTK contributions $\sigma_{tot} = \alpha\sigma_1 + (1-\alpha)\sigma_2$, with parameters $\Delta_1$, $\Gamma_1$, $Z_1$, $\Delta_2$, $\Gamma_2$, $Z_2$ and the respective weights $\alpha$, and (1-$\alpha$). This model, which we used as the first to describe the two-band PC spectra in $MgB_2$ [11], has been successfully used to describe PC data in other multi-gap superconductors as pnictides [12,13], borocarbides [14], and in noncentrosymmetric superconductors $Re_6Zr$ [15] and BiPd [16].

Heat capacity measurements have been performed using an ac technique as described elsewhere [17,18]. In the experiment, heat was supplied by a light-emitting diode and was driven to the sample using an optical fiber. The temperature oscillations of the sample were recorded by a thermocouple that served as a sample holder at the same time. Such configuration reduces significantly the addenda contribution to the total measured heat capacity and enables measurement of the sample's heat capacity with high resolution. However, the resulting data is in arbitrary units.

Local magnetometry measurements using miniature Hall probes were realized in small magnetic fields and temperatures down to 0.5 K. An array of several Hall probes arranged in a line was employed to detect local magnetic induction in the sample. The dimensions of each probe were 10 x 10 $\mu m^2$, with the spacing between two adjacent probe centers 35 $\mu m$. The sample fragment, cut from the sample that was used for PCAR measurements, was placed on top of the array, secured by a small drop of vacuum grease and cooled in zero magnetic field. The array was powered by a current of 20 $\mu A$ and a voltage across each probe was recorded in a gradually increasing or decreasing magnetic field. The measured voltage across the probe is proportional to the magnetic induction at the position of the probe. Details of the method are described e.g. in Ref. [19] and partially in the Suppl. Mat. [20].

## 3 Results

PCAR measurements were performed on several $La_3Se_4$ samples using different PC needles (Au, Cu, Pt). We found that ballistic PC's can most reliably be formed with a tip sharpened from a 100 $\mu m$ Pt wire at PC resistances $R_{PC}$ between 7 $\Omega$ and 60 $\Omega$. Most of the measured spectra showed two symmetric peaks near ±2 meV, a typical manifestation of the Andreev reflection on PC's with the presence of a contact barrier (Fig. 1, curve 1) indicating a single superconducting energy gap. The spectrum was fitted to the single-gap BTK model, with parameters $\Delta = 1.9$ meV, $\Gamma = 1.1$ meV, $Z = 0.6$ as indicated by the blue dashed curve in Fig. 1.

In some cases, usually on spectra with higher contact-barrier strength, we observed an additional maximum at zero bias (Fig. 1, curve 2) or a weakly split structure near zero bias (curve 3) along the dominant maxima, similar to those in spectrum 1. The single-gap model clearly cannot describe these spectra. The PCAR spectra 2 and 3 were thus fitted to the two-gap BTK formula (red dotted lines). We can see that this model describes our spectra (solid lines) reasonably with the parameters of two energy gaps $\Delta_1 = 2.2$ meV, $\Delta_2 = 0.75$ meV, and the spectral weight $\alpha = 0.85$ for the spectrum 2 and $\Delta_1 = 2.2$ meV, $\Delta_2 = 0.9$ meV, and the spectral weight $\alpha = 0.85$ for the spectrum 3. The other relevant parameters are listed in the figure caption.

Accounting on the statistics of another 10 spectra we found the averaged values of the energy gap $\Delta_1 = 2.05 \pm 0.15$ meV, $\Delta_2 = 0.85 \pm 0.15$ meV and spectral weight $\alpha$ varying between 0.85 and 0.9 for the large gap at $T_c = 8.1 \pm 0.3$ K, providing the superconducting gap ratios $2\Delta_1/k_BT_c \sim 5.8$ and $2\Delta_2/k_BT_c = 2.3$.

Obviously, it is difficult to directly identify the second energy gap with a weight of about 10% in the PCAR spectra. Favourable conditions occur when the large-gap spectrum has a significant tunneling character, with $Z > 0.5$, forming a dip between the gap-like peaks where the small gap can appear. Also, different values of the spectral smearings at individual gaps can help in the direct observation of the two-gap structure. For example, in Fig. 1 we can see that with increasing smearing at the large gap the structure of the small gap becomes more pronounced.

The spectrum 2 in Fig. 1 reveals a central maximum which can be fitted considering the small gap with $\Delta_2 = 0.75$ meV and the corresponding weight $(1-\alpha) = 0.15$. Small $Z_2 = 0.1$, due to the high transparency of the junction for the electrons from the related band, causes the gap to appear as a maximum, not a pair of symmetric peaks as for the large gap.

Yet, there are another possibilities to explain such behavior. The observed structure could be the zero-bias conductance peak (ZBCP) related to the unconventional order parameter. Recently, He and Chen [21] analyzed a possible triplet superconductivity with an Anderson-Brinkman-Morel state in a PCAR spectra of Bi/Ni bilayer measured by Zhao et al. [22]. They found that the PCAR conductance spectra always show a ZBCP when an incident quasiparticle from the superconducting side will change sign after it is reflected by the interface in the point contact. Since our $La_3Se_4$ sample is polycrystalline we do not control the direction of the PC current relative to the crystal structure. On the other hand with a relatively small antisymmetric spin-orbit splitting energy 65 meV [7], the probability of unconventional behavior is not very high. The splitting energy is close to that found in another La-based noncentrosymmetric superconductor, $LaNiC_2$, where two-gap superconductivity is observed [6].

In the following we focus on a possibilty of two-gap superconductivity in the system and we study the effect of temperature and magnetic field on the PCAR spectra. This required the formation of a stable microconstriction capable to withstand the time-consuming cycling of temperatures and magnetic fields. Figures 2 and 3 show the effect of temperature and magnetic field, respectively, on the microconstriction displayed as the curve 1 in Fig. 1.

In Figure 2 the apparent single-gap structure of the spectrum (solid line) is gradually suppressed with increasing temperature up to $T \sim 4.5$ K as expected in the case of a single-gap superconductor, but at higher temperatures, features of a second, smaller, gap appear. It is clearly visible that besides the shrinking and broadening of two maxima, a central peak appears at $T = 5.5$ K. This central peak is even more pronounced at 6 and 6.5 K and can indicate the contribution of the small gap smeared by temperature. At higher temperatures, this triple structure gradually smears out into a single maximum that closes above $T \sim 8$ K. The dotted lines in Fig. 2 show our fitting curves to the two-gap BTK model at all temperatures from 2.2 K to 8 K. Since the spectra at the lowest temperatures do not show any apparent two-gap structure, we started the fit at $T = 5.5$ K, where the two gaps are resolved. We found that this spectrum can be described in good agreement with the two-gap BTK model with energy gaps $\Delta_1(5.5\text{ K}) = 1.75$ meV, $\Delta_2(5.5\text{ K}) = 0.78$ meV and parameters $Z_1 = 0.75$, $\Gamma_1 = 1.15$ meV, $Z_2 = 0.1$, $\Gamma_2 = 0.1$ meV and weight $\alpha = 0.85$. We used these $Z$, $\Gamma$ and $\alpha$ parameters as the initial-fit values at all temperatures from $T = 2.2$ K to 8 K allowing for corrections of about 10%.

The main results of the fitting procedure, the temperature dependencies of the two energy gaps, are shown in the inset of Fig. 2 with open symbols. It is clear that both gaps close at the same critical temperature near 8 K. The thick gray and red lines in the inset represent the BCS-like $\Delta(T)$ dependencies with $\Delta_1(0) = 2$ meV, $\Delta_2(0) = 0.8$ meV and $T_c = 8$ K defining the gap ratios as $2\Delta_1/k_BT_c \sim 5.8$ and $2\Delta_2/k_BT_c \sim 2.3$.

The effect of the magnetic field on selected PCAR spectra from Fig. 2 has been measured at six different temperatures. The dependencies at temperatures $T = 2.2$ K, 5 K, and 7.5 K are summarized in Figure 3. The magnetic field evolution measured at $T = 2.2$ K (a) shows a classical suppression of the superconducting gap structure up to $B = 4$ T, i.e. with increasing magnetic field the gap is smeared out and the positions of the symmetrical gap maxima are shifted to lower energies. However, at $B = 5$ T an additional structure appears at the zero bias, forming a three-hump maximum. This structure is smearing out up to $B = 8$ T, where only a central maximum is left. The two-gap structure of the spectrum is clearly visible at $T = 5$ K already in the zero magnetic field (Fig. 3 (b)). It becomes smeared into a single maximum at $B \sim 2$ T and completely disappears at a magnetic field of 8 T, which defines the value of the upper critical magnetic field at a given temperature $B_{c2}(5\text{ K})$. Figure 3 (c) shows the magnetic field dependence of the PCAR spectra at $T = 7.5$ K. Although this temperature is close to $T_c$ and causes significant thermal smearing, we can see the two-gap structure even here. From the smeared maximum at $B = 0$ T, the three-hump structure develops with the increasing magnetic field, visible already at a small field of 0.1 T. The structure then gradually washes out and disappears completely at the critical magnetic field $B_{c2}(7.5\text{ K}) = 1.8$ T.

Next, from our PCAR spectroscopy measurements in various magnetic fields, we determined the upper critical field $B_{c2}$, as the lowest field where no superconducting spectral feature is present. Such an approach, sensitive to the superconducting order parameter, is closer to the thermodynamic quantity of $B_{c2}$ than usual transport measurements, which can be affected by the vortex dynamics [23,24]. Figure 4 shows the temperature dependence of $B_{c2}(T)$ determined from PCAR measurements (solid symbols). The red line shows the classical Werthamer-Helfand-Hohenberg (WHH) curve [25] generated for a one-gap superconductor with $T_c = 8.3$ K and $B_{c2}(0) = 14$ T. In contrast to the resistive determination of $B_{c2}$ published in the previous paper [7] we do not see any positive curvature of $B_{c2}$ close to $T_c$. It is generally believed that a positive curvature of the temperature dependence of $B_{c2}$ in the vicinity of $T_c$ is a manifestation of the two-gap superconductivity. This was also considered in our previous paper [7] where the resistively determined $B_{c2}$ shows a positive curvature at high temperatures near $T_c$ similarly to the iconic case of MgB$_2$ [26, 27]. The shape of the temperature dependence of the $B_{c2}$ curves in two-gap superconductors varies considerably depending on the ratio of the diffusion coefficients in the individual bands and on the values of intra- and inter-band couplings [28-30]. As a consequence, not only a positive curvature, but also a linear dependence of $B_{c2}(T)$ can be observed in the vicinity of $T_c$ in several cases [28-30].

Point-contact spectroscopy is a surface-sensitive method. In order to inspect the bulk properties of the sample we also performed the heat capacity measurements by a very sensitive ac calorimetry. Figure 5 displays the total heat capacity divided by temperature (symbols) measured below 10 K in zero magnetic field. Note that only one out of 30 points is displayed as a symbol for clarity. Below 7.4 K, a relatively broad anomaly appears in the data, related to a transition to the superconducting state. Smearing of the anomaly reflects the polycrystalline character of the sample and a possible existence of several close phases. The superconducting transition is then followed by an exponential decrease of the data with decreasing temperature. At low temperatures, another anomaly was detected, with the mid-point at 2.4 K. The origin of this anomaly will be discussed below.

Due to the high value of the upper critical magnetic field of the sample (see Fig.4 and [7]) it was not possible in our set-up to suppress superconductivity entirely to measure the normal-state contribution to the heat capacity in the overall temperature range directly. Instead, the normal-state heat capacity has been obtained based on the zero-field measurement above the superconducting transition. For the fit, a standard formula for the low-temperature normal-state heat capacity $C_n$ was used, consisting of electronic and phononic contributions in the form $C_n = aT + bT^3 + cT^5$. The result of the fit extrapolated down to low temperatures is depicted in the figure as the dashed line. A phenomenological model – the so-called alpha model for a two-gap superconductor [31] was applied to analyze the heat capacity data. In the model, the fitting parameters are the ratios of $\Delta/k_BT_c$ corresponding to each of the two energy gaps and their respective weight ($\alpha$ for the large, and (1-$\alpha$) for the smaller energy gap). Due to the pronounced smearing of the main anomaly and the existence of the low-$T$ anomaly, from a comparison of the data with the model we obtained quite large distribution of resulting parameters: $2\Delta_1/k_BT_c = 5.75 \pm 0.25$ for the larger energy gap, $2\Delta_2/k_BT_c = 2 \pm 0.5$ for the smaller one, and $\alpha = 0.8 - 1$. In Fig. 5 the model curve with $2\Delta_1/k_BT_c = 5.75$, $2\Delta_2/k_BT_c = 2$, and $\alpha = 0.9$ is displayed by the red line. Apart from the two regions around anomalies, the model curve follows the data in a fair agreement.

To address the low-temperature anomaly, we performed additional heat capacity measurements in the low-temperature region, in magnetic fields up to 8 T (see Fig. S1 in the Suppl. Mat. [20]). Increasing magnetic field gradually suppresses the height of the anomaly while its position seems to be affected very weakly. If the anomaly were related to a superconducting transition, the upper critical magnetic field could be above 50 T (see the inset of Fig. S1 in Suppl. Mat. [20]). We consider this very unlikely.

To make another test we perform local magnetometry measurements by the Hall probes. As shown below they also testify the absence of the superconducting phase with a low critical temperature around 2 K. The Hall probes are able to detect the so-called penetration field ($B_p$) at which the superconducting sample quits the Meissner state and lets the magnetic field pierce through it. The penetration field is, by a geometrical factor, related to the lower critical magnetic field, which is proportional to the superfluid density. Then, the temperature dependence of $B_p(T)$ bears information about the ratio $\Delta/k_BT_c$. Figure 6 shows the resulting data of $B_p(T)$, derived from the measurements on one of the probes – symbols (procedure to determine $B_p$ is described in the text and Fig.S2 in the Suppl. Mat. [20]). For our purpose, the low-temperature part is important - if there were a superconducting phase in the sample with the $T_c$ around 2.4 K, the data would show an abrupt increase with the decreasing temperature below that temperature (see Fig. S3 in Suppl. Mat. [20]). In $B_p(T)$ we did not observe anything like that, instead, the data saturates. Moreover, a theoretical curve (red dashed line in Fig.6) involving two energy gaps, with the same parameters as those obtained from the heat capacity measurements, follows our data in a very good agreement proving the two gap superconductivity in the system. We checked several Hall probes distant at ~ 100 μm and arrived to very similar temperature dependence of $B_p$. It is also noteworthy that the Hall probe magnetometry and the ac calorimetry have been done on the same piece of sample cut from the La$_3$Se$_4$ polycrystal used for PCAR measurements.

The findings that the position of the low-temperature anomaly observed in the heat capacity is almost independent on the applied magnetic field (Fig. S1 in Suppl. Mat. [20]) and the temperature dependence of the penetration magnetic field $H_p$ does not show the presence of a low-temperature phase below $T = 3$ K (Fig. 6) rule out the superconducting origin of the low-temperature anomaly in heat capacity. The source of this anomaly is probably related to the presence of local magnetic moments at the grain boundaries of the disordered polycrystalline sample, where uncompensated electron orbitals on the "dangling bonds" may be the source of local magnetism [32]. Yet, a presence of a secondary magnetic phase cannot be excluded. A similar low-temperature anomaly has been observed in FeSe samples [33], where it's presence has been explained with a possible antiferromagnetic order.

Preliminary measurements of the specific heat with the physical property measurement system (PPMS, Quantum Design) above 5 K [7] showed a superconducting transition at around 7.7 K and a jump pointing to a strong coupling with $\Delta C/\gamma T_c$ to be 2 with the Sommerfeld coefficient $\gamma = 25.7$ mJ mol$^{-1}$ K$^{-2}$. Our recent ac calorimetry measurements down to 1.5 K confirmed that the system reveals a strong superconducting coupling. Moreover, the two-gap alpha model can fit the superconducting part consistently with the results of the point-contact Andreev reflection, as well as of the Hall-probe magnetometry.

## 4 Discussion

The results of our PCAR measurements clearly demonstrate the existence of multi-band superconductivity in our system. We have shown the existence of two energy gaps not only from the fitting of the temperature dependence of the PCAR spectra (Fig. 2), but also from direct PCAR measurements, where due to the different sensitivity of the individual gaps to the applied magnetic field we have seen the presence of the two-gap structure even in close proximity to the critical temperature of the bulk sample (Fig. 3). It is important to notice, that we have used this method successfully also some-times ago for the identification of multiband superconductivity in MgB$_2$ [11]. Being able to determine the temperature dependence of the both gaps in our samples also gives us important information about the nature of the two-gap superconductivity. Already in their pioneering work Suhl *et. al* [34] in 1959 showed that in systems with overlapping bands two-gap superconductivity can occur, where due to inter-band interactions two different superconducting gaps can close at the same $T_c$. In systems with a weak interaction between the separated bands the small gap $\Delta_2$ should show a significant deviation from the BCS dependence in the form of tailing at temperatures increasing towards $T_c$. The fact that both our $\Delta_{1,2}(T)$ dependencies follow the BCS dependence indicates, that there is a strong inter-band interaction in our system. The presence of strong inter-band interaction is also indicated by the shape of the temperature dependence of $B_{c2}$, shown in Fig. 4. As mentioned above, the temperature dependence of $B_{c2}$ in two-gap superconductors does not necessarily exhibit a positive curvature in the vicinity of $T_c$. The change of curvature arises due to different values of diffusion coefficients in the individual bands, however, the increase of inter-band interaction leads to a flattening of the positive curvature into a linear dependence in the vicinity of $T_c$ [27-30].

Our PCAR results are also strongly supported by our heat capacity and magnetization measurements. The fact that we can describe the temperature dependence of the heat capacity by the two-gap alpha model with $2\Delta_1/k_BT_c = 5.75$, $2\Delta_2/k_BT_c = 2$, and $\alpha = 0.9$ (Fig. 4), and that the same parameters can also describe the temperature dependence of the penetration field $B_p$, determined from Hall probe magnetometry indicates that the observation of two gaps with the same $T_c$ is not only a surface effect, but it is a bulk property of the La$_3$Se$_4$ samples.

## 5 Conclusions

To conclude, we have performed the PCAR studies of the order parameter in the nocentrosymmetric superconductor of La$_3$Se$_4$ for the first time. Although on the majority of the spectra only a single gap at about 2 meV is directly observed on some junctions another smaller gap appears. Most importantly, at higher temperatures and in magnetic fields the second, smaller gap emerges also in the seemingly single-gap spectra due to a different sensitivity to the magnetic field. Then, the both gaps can be followed up to the common critical temperature of the bulk sample. Highly sensitive ac calorimetry and Hall-probe magnetometry measurements support the two-gap superconductivity consistently with PCAR measurements.

**Funding Information** This work was supported by Projects APVV-20–0425, VEGA 2/0073/24, COST Action No. CA21144 (SUPERQUMAP), Slovak Academy of Sciences Project IMPULZ IM-2021–42. The work in Dresden was supported by the German Research Foundation (DFG) within the Walter Benjamin research grant NA 2012/1-1.

# Figures

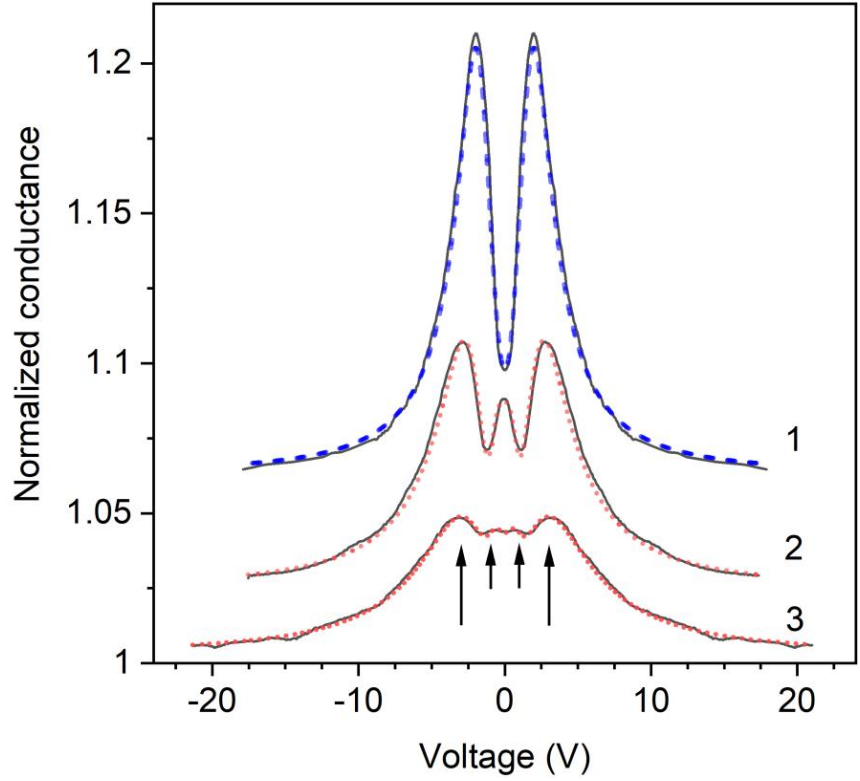

**Fig. 1**: Typical PCAR spectra measured in Pt-La$_3$Se$_4$ point contacts (solid black lines) at *T* = 2 K (curve 1) and 1.5 K (curves 2, and 3). Spectrum 1 is fitted to the one-gap BTK model, with parameters Δ = 1.9 meV, Γ = 1.1 meV, Z = 0.6 (blue dashed curve). Curve 2 and 3 are fitted to the two-gap BTK model with the parameters for spectrum 2 - Δ$_1$ = 2.2 meV, Γ$_1$ = 1.72 meV, Z$_1$ = 0.7, Δ$_2$ = 0.75 meV, Γ$_2$ = 0.23 meV, Z$_2$ = 0.1, α = 0.85 and spectrum 3 - Δ$_1$ = 2.2 meV, Γ$_1$ = 2.6 meV, Z$_1$ = 0.65, Δ$_2$ = 0.9 meV, Γ$_2$ = 0.51 meV, Z$_2$ = 0.3, α = 0.85 (red dotted curves). The black arrows mark the gap positions at their average values. The spectra are shifted around the Y-axis for clarity.

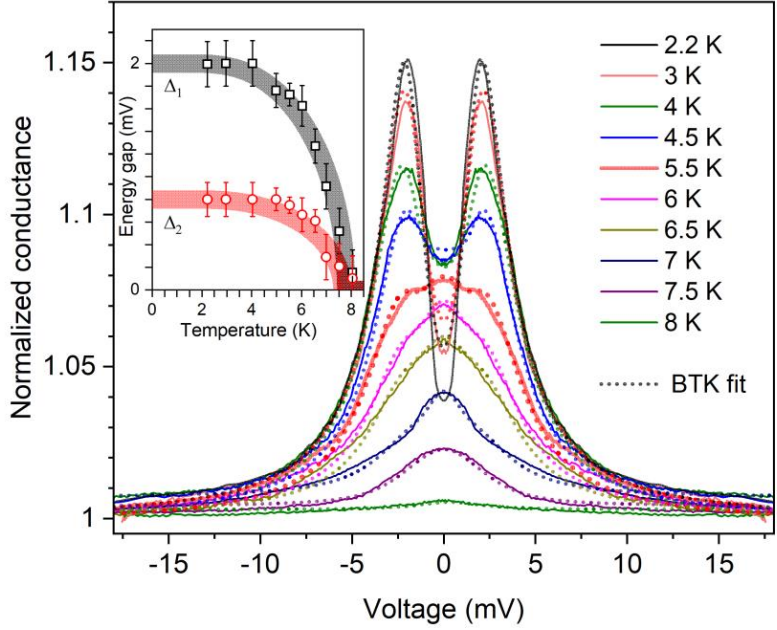

**Fig. 2:** Temperature dependence of the PCAR spectra measured at indicated temperatures (solid lines). The dashed lines are the fitting curves to the two-gap BTK formula with $Z_1 = 0.7 \pm 0.1$, $\Gamma_1 = 1.15 \pm 0.1$ meV, $Z_2 = 0.1 \pm 0.05$, $\Gamma_2 = 0.15 \pm 0.05$ meV, weight $\alpha = 0.85 \pm 0.1$ and energy gap values shown in the inset with open symbols. The bold gray and red curves in the inset are BCS-like $\Delta(T)$ dependencies with $\Delta_1(0) = 2$ meV, $\Delta_2(0) = 0.8$ meV, and $T_c = 8$ K, giving the strengths of the superconducting coupling $2\Delta_1/k_B T_c \sim 5.8$ and as $2\Delta_2/k_B T_c \sim 2.3$.

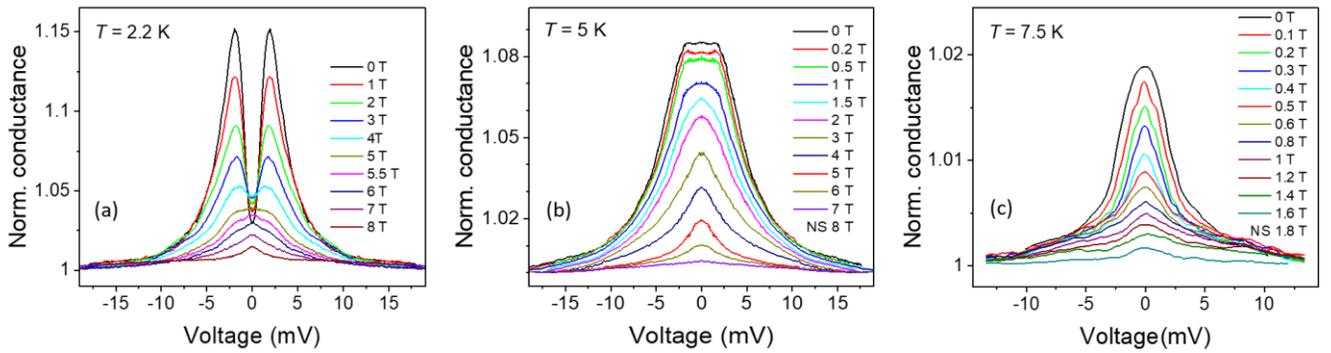

**Fig. 3:** Magnetic field dependencies of PCAR spectra from Fig. 2 at selected fixed temperatures. The two-gap features appear around $B = 5.5$ T at $T = 2.2$ K and low fields at $T = 5$ K and $T = 7.5$ K, resp.

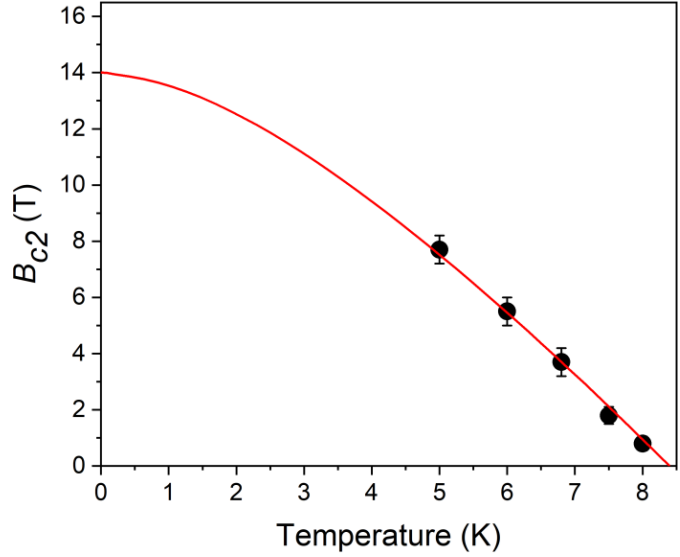

**Fig. 4**: Temperature dependence of the upper critical magnetic field $B_{c2}(T)$ determined from the PCAR measurements performed at fixed temperatures and different magnetic fields. The normal state transition was defined at the field, where the superconducting gap structure of the PCAR spectra disappeared. The red line is the fit to WHH theory, which defines $B_{c2}(0) = 14$ T.

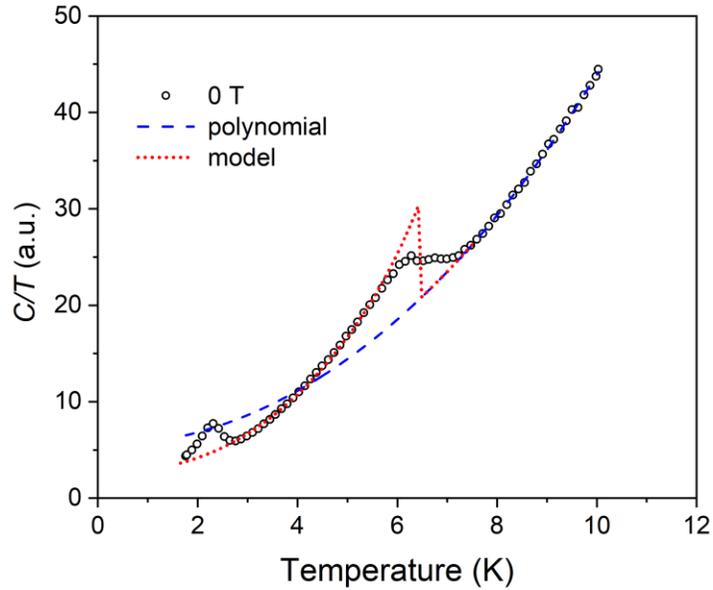

**Fig. 5**: Total heat capacity measured in zero magnetic field (black circles, only one out of 30 points is shown for clarity), a polynomial fit of the heat capacity above the superconducting transition i.e. in the temperature range 7.4-10 K (blue dashed line), alpha model (red dotted curve) with $2\Delta_1/k_B T_c = 5.75$, $2\Delta_2/k_B T_c = 2$, and $\alpha = 0.9$

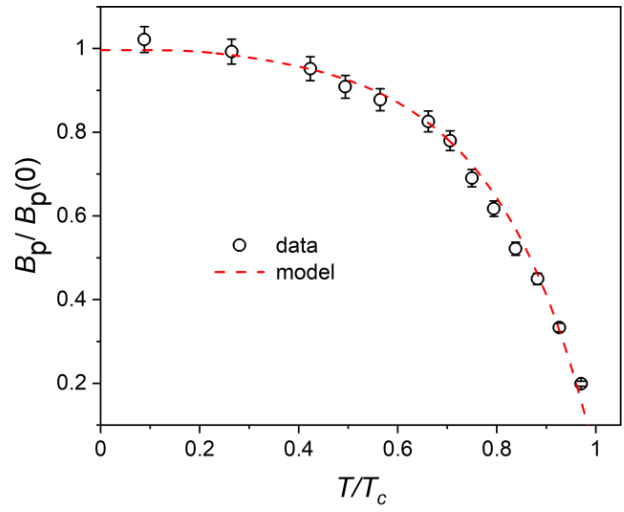

**Fig.6**: Temperature dependence of the penetration field (symbols) proportional to the superfluid density. The theoretical curve of the superfluid density, calculated as a weighted sum of two contributions with the same parameters as in Fig.4 ($2\Delta_1/k_BT_c =$ 5.75 with $\alpha = 0.9$, and $2\Delta_2/k_BT_c = 2$, with $1-\alpha = 0.1$), rescaled to match the data (red dashed line).

# Supplemental Material

## Two-gap superconductivity in the noncentrosymmetric La$_3$Se$_4$ compound

F. Košuth,[1,2] N. Potomová,[2] Z. Pribulová,[1] J. Kačmarčík,[1] M. Naskar,[3] D.S. Inosov,[3] S. Ash,[4] A.K. Ganguli,[5,6] J. Šoltýs,[7] V. Cambel,[7] P. Szabó,[1] and P. Samuely[1]

[1]Centre of Low Temperature Physics, Institute of Experimental Physics, Slovak Academy of Sciences, SK-04001 Košice, Slovakia
[2]Centre of Low Temperature Physics, Faculty of Science, P. J. Šafárik University, SK-04001 Košice, Slovakia
[3]Institut für Festkörper- und Materialphysik, Technische Universität Dresden, D-01069 Dresden, Germany
[4]Institute for Solid State Research, Leibniz IFW Dresden, 01069 Dresden, Germany
[5]Department of Chemistry, Indian Institute of Technology Delhi, New Delhi 110016, India
[6]Department of Chem. Sci., Indian Institute of Science Education and Research Berhampur, Odisha 760003, India
[7]Institute of Electrical Engineering, Slovak Academy of Sciences, Bratislava, Slovakia

**Low temperature heat capacity measurements**

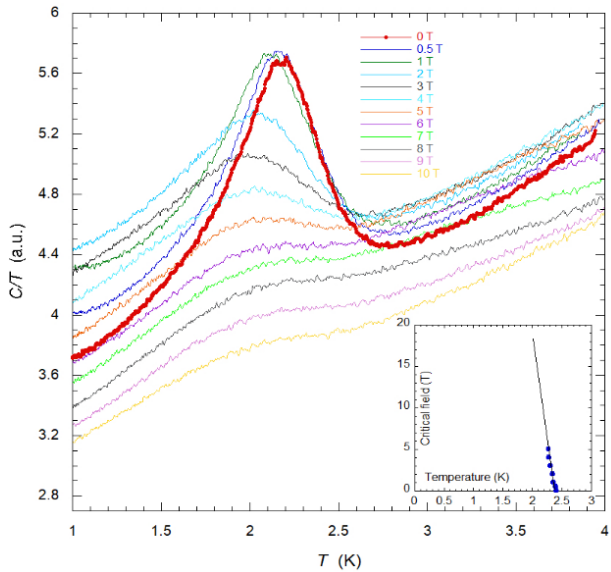

Figure S1: Temperature dependence of the low temperature heat capacity anomaly measured at temperatures 0.8 K – 4 K and magnetic fields indicated in the figure. The inset shows temperature dependence of the midpoint of the heat capacity transition at different magnetic fields (blue symbols). The black line is the linear extrapolation of the blue symbols indicating an extremely high value of the critical field at T → 0.

**Determination of the penetration field**

Prior to each Hall probe measurement at fixed temperature, the sample was first heated above the critical temperature and subsequently cooled in zero magnetic field to ensure the absence of vortices being trapped in the sample. A small current was necessary to supply to the superconducting magnet to compensate for the remanent field of the coil. The measurements were realized first in gradualy increasing magnetic field up to 15 mT and then decreasing back to 0 T. When the sample is in an ideal diamagnetic state, the magnetic field does not penetrate the sample. The probes are shielded from the field by the sample so that no Hall voltage is detected. When the magnetic field penetrates the sample, it is reflected in an increase of the magnetic induction read by the Hall probe. However, due to non-zero distance between the probe and the sample, the Hall probe picks some portion of the external field even when the sample is in Meissner state. This is observed as an initial linear increase (see left panel of Fig. S3), which is removed before further data treatment (middle panel of Fig. S3). Due to presence of some pinning centers in the sample, the increase of the voltage after the field penetration has a shallow, quadratic form at low fields. To have a clear criterion to determine the field where the signal starts to deviate from zero, square root of the data (after calculation of the absolute value of the data) was calculated. The penetration field $B_p$ was then assigned to the value where the two lines on the right panel intercept. The same procedure was performed with the other measurements at different temperatures as well.

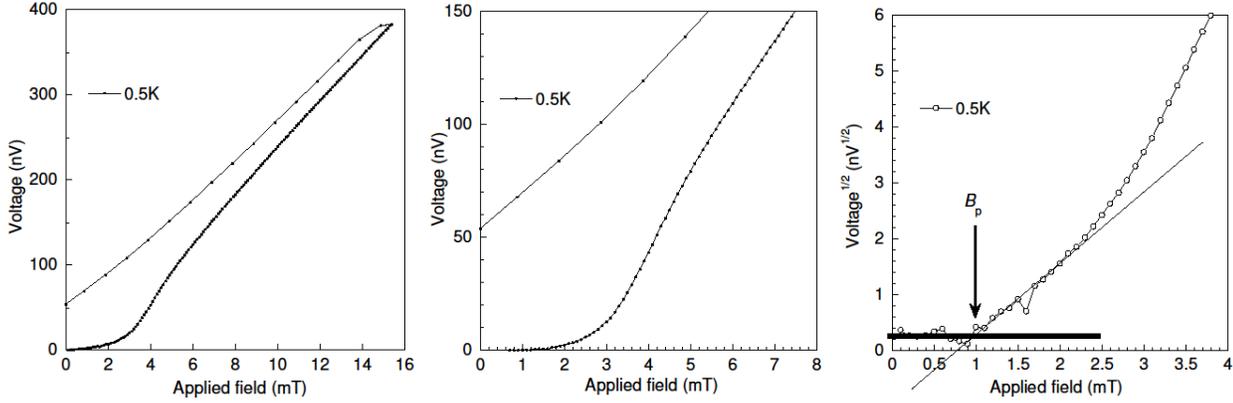

Figure S2: Procedure to determine the penetration field: Left panel – raw data at 0.5 K; Middle panel – data after subtraction of the initial linear increase; Right panel – square root of the data, the lines are guide to the eyes – penetration field is determined at the magnetic field where the two lines intercept.

**Model curves for the superfluid density**

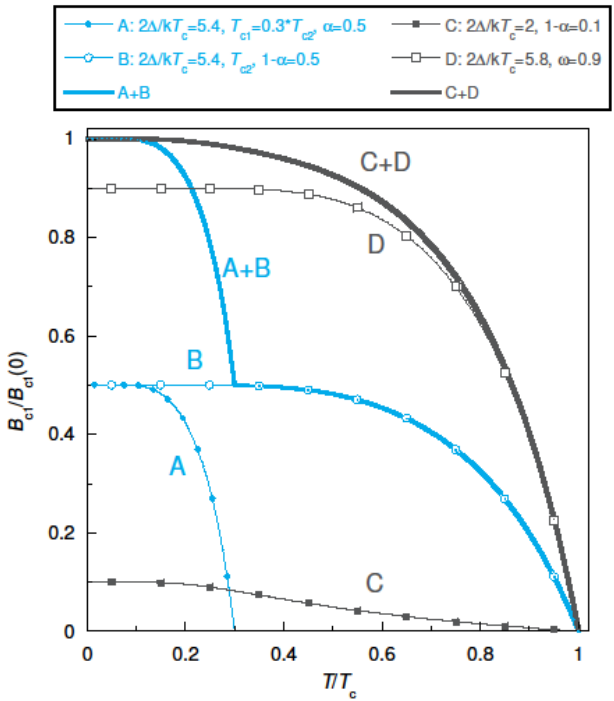

Figure S3: model curves of the superfluid density
$\frac{B_{c1}(T)}{B_{c1}(0)} \approx \frac{\lambda^2(0)}{\lambda^2(T)} = 1 - 2\int_{\Delta(T)}^{\infty} \frac{\partial f}{\partial E} \frac{E}{\sqrt{E^2 - \Delta^2(T)}} dE$, where $B_{c1}$ is the lower critical magnetic field, $\lambda$ is the penetration depth and $\Delta$ is the energy gap – thick lines are weigted sum of two contributions for the case of two superconducting phases with different critical temperature (A+B in blue), and for the case of two energy gaps with different ratio $\Delta/k_B T_c$ closing at the same $T_c$ (C+D in grey). The grey curve (C+D) is used in Fig.5 of the article to model the temperature dependence of our data.